\newif\ifcomments  
\newif\ifsupp  
\commentstrue
\supptrue

\documentclass[a4paper]{article}

\usepackage{INTERSPEECH2022}
\usepackage{soul}

\title{Extracting Targeted Training Data from ASR Models, \\
and How to Mitigate It}
\name{Ehsan Amid$^*$\thanks{$^*$Equal contribution.}, Om Thakkar$^*$, Arun Narayanan, Rajiv Mathews, Françoise Beaufays}

\address{Google LLC}
\email{\{eamid, omthkkr, arunnt, mathews, fsb\}@google.com}

\usepackage{amsmath,amssymb,pifont}
\usepackage{multicol}
\usepackage{amstext}
\usepackage{amsthm}
\usepackage{multirow}
\usepackage{booktabs}
\usepackage[skip=0pt]{subcaption}
\usepackage{times}
\usepackage{lipsum}
\usepackage[shortlabels]{enumitem}
\usepackage{cancel}
\usepackage{wrapfig}
\usepackage{array}
\usepackage{siunitx}
\usepackage{csvsimple}
\usepackage[multidot]{grffile}
\usepackage{bbm}
\usepackage{dblfloatfix}
\usepackage{hyperref}
\usepackage{makecell}
\usepackage{bbm, dsfont}
\usepackage{mathtools}
\usepackage{xcolor}
\usepackage{comment}
\usepackage{diagbox}
\usepackage[multiple]{footmisc}
\usepackage{mathrsfs}
\usepackage{tikz}
\usepackage{hyperref}
\usepackage{cleveref}
\usepackage{algpseudocode,algorithm,algorithmicx}
\usepackage{xfrac}

\hypersetup{final}

\newcommand{\attname}{Noise Masking}

\newcommand{\augname}{Word Dropout}

\newcommand{\uppname}{Name Silencing}

\renewcommand{\vec}[1]{#1}

\makeatletter
\newcommand{\vast}{\bBigg@{4}}
\newcommand{\Vast}{\bBigg@{5}}
\makeatother

\newcommand{\ex}[2]{{\ifx&#1& \mathbb{E} \else
\underset{#1}{\mathbb{E}} \fi \left[#2\right]}}
\newcommand{\pr}[2]{{\ifx&#1& \mathbb{P} \else
\underset{#1}{\mathbb{P}} \fi \left[#2\right]}}

\DeclarePairedDelimiterX{\ip}[2]{\langle}{\rangle}{#1, #2}

\DeclarePairedDelimiterX{\infdivx}[2]{(}{)}{%
  #1\;\delimsize\|\;#2%
}

\newcommand{\mypar}[1]{\smallskip
	\noindent{\textbf{{#1}:}}}
	
\renewcommand{\epsilon}{\varepsilon}

\usepackage{tikz}
\usepackage{color}

\begin{document}

\maketitle
\begin{abstract}
Recent work has designed methods to demonstrate that model updates in ASR training can leak potentially sensitive attributes of the utterances used in computing the updates. 
In this work, we design the first method to demonstrate information leakage about training data from trained ASR models.
We design \attname{}, a fill-in-the-blank style method for extracting targeted parts of training data from trained ASR models.
We demonstrate the success of \attname{} by using it in four settings for extracting names from the LibriSpeech dataset used for training a state-of-the-art Conformer model.
In particular, we show that we are able to extract the correct names from masked training utterances with 11.8\% accuracy, while the model outputs some name from the train set 55.2\% of the time. 
Further, we show that even in a setting that uses synthetic audio and partial transcripts from the test set, our method achieves 2.5\% correct name accuracy (47.7\% any name success rate).
Lastly, we design \augname{}, a data augmentation method that we show when used in training along with Multistyle TRaining (MTR), provides comparable utility as the baseline, along with significantly mitigating extraction via \attname{} across the four evaluated settings.
\end{abstract}
\noindent\textbf{Index Terms}: Information extraction, Data augmentation
\section{Introduction}
\label{sec:intro}

Modern end-to-end Automatic Speech Recognition (ASR) systems are increasingly being trained to improve inference in the presence of background noise (e.g., \cite{MTR, NMSPTEH18, spec}). 
Training such models often requires audio and transcripts of millions of utterances.
Recent work has designed methods to demonstrate that model updates in ASR training can leak potentially sensitive attributes like labels~\cite{DTR21} and speaker identity~\cite{DTR221} of utterances used in computing the updates.
Many prior works have also focused on demonstrating that trained language models (LMs) \cite{CLEKS19, SS19, TRMB20, RTM20, CTWJHLRBS21, CIJLTZ} and image classification models \cite{FZ20, BCH} are susceptible to \emph{unintended memorization} of the rare or unique data that was part of the train set. 
Even models trained on large scale datasets can leak potentially sensitive information about their training data, and  the above-listed works design various methods to demonstrate such leakage.

There can be various situations where extracting targeted parts of training data (e.g., specific words following a general structure) can leak the participants' privacy. For instance, obtaining meaningful extractions from a model for phrases like ``Alice is infected with [...] disease", or ``Bob is [...] years old" can leak sensitive information about the samples used for training it.
In this work, we study if targeted parts of 
training data can be extracted from ASR models  using only a query-access to the model.
Note that any method requiring only a black-box access has wide applicability since it can operate on even deployed ASR models to showcase privacy leakages.
We design a fill-in-the-blank style method called \attname{} that can demonstrate the susceptibility of  ASR models to leak memorized parts of their training data using only query-access to them.
To our knowledge, \attname{} is the first method to showcase leakage about training data from trained ASR models.

We evaluate our \attname{} method by performing experiments on a state-of-the-art Conformer (L) model \cite{gulati2020conformer} trained on the benchmark LibriSpeech dataset \cite{PCPK15}.
Leveraging the fact that LibriSpeech is a corpus of read English audiobooks, and some  minimal knowledge about the language in English books, we show that using \attname{} on a trained Conformer model can leak memorized names from the dataset.

We also design a data augmentation method, \augname{}, for training ASR models that we empirically show to provide robustness against \attname{} when used for training with Multistyle TRaining (MTR)~\cite{MTR}.
Further, we also show that training with \augname{} provides comparable utility as a baseline not using \augname{}.

\mypar{Organization of the paper} In Section~\ref{sec:nm}, we describe the details of our \attname{} method, and provide an empirical evaluation of it.
Next, in Section~\ref{sec:mitigs}, we provide an evaluation of various training methods, including our \augname{} method, towards mitigating \attname{}.
We show some detailed results for practitioners to analyze data leakage in Section~\ref{sec:dets}. 
We state the conclusions of this work in Section~\ref{sec:conc}.

\section{Extraction via Noise Masking}
\label{sec:nm}

\subsection{The \attname{} Method}
\label{sec:method}
We provide a general description of the \attname{} method for extracting information about the training data given an ASR model.
Assume that an ASR model $M$ is trained over a dataset $D$ consisting of training examples from some population $\mathcal{P}$, and query-access to $M$ is made available.
At its most basic level, to design a query via \attname{}, an analyst requires:
\begin{enumerate}[leftmargin=*]
    \item $I(\mathcal{P})$, which denotes some target knowledge regarding the population. This knowledge can take various forms;  e.g., a street name typically appears before the word ``Street", or a number typically follows the word ``Apartment", etc. The analyst may guess the existence of such target structures in the training data and move on to the next step.
    \item $T$, a procedure for obtaining relevant transcript(s) given some target knowledge about a population. For example, with target knowledge about addresses, $T$ could provide transcripts like ``Alice lives in Apartment [...] on [...] Street".
    \item $U$, a procedure for generating utterances given transcripts. For instance, a Text-To-Speech (TTS) system.
    \item $N$, a procedure for incorporating noise into given utterances. For example, an audio of some music segment.
\end{enumerate}

The analyst can use the target knowledge $I(\mathcal{P})$ to obtain a candidate transcript $t = T(I(\mathcal{P}))$. This can be subsequently used to generate an utterance $u = U(t)$, and incorporate noise into it as $q = N(u)$ for querying the model $M$ with $q$ and obtaining an output $t'$.
The analyst can compare $t$ and $t'$ to evaluate for extractions from the training data $D$.

In many scenarios, it can be easy for an analyst to guess target knowledge based on the intended use-case(s) of an ASR model.
For successful extractions, it can be crucial to get relevant transcripts with such knowledge.
However, our method requires only a query-access to a trained model, which allows adaptive applications of it in many common settings.

\subsection{Empirical Evaluation}
\label{sec:expts}
Now, we provide an instantiation of our \attname{} method, and conduct experiments to evaluate its performance. 

\mypar{Train Dataset} We use the  LibriSpeech dataset\footnote{Available online at: \url{https://www.tensorflow.org/datasets/catalog/librispeech}}~\cite{PCPK15}, which contains $\sim$\num{1000} hours of English audiobook recordings by several speakers. 
The dataset contains numerous names from books, 
with a lot of them appearing after \emph{title words} like `mister', `miss', `missus',
etc.
For our experiments, we consider such names as an example of \emph{sensitive} information: suppose using \attname{}, a trained model reveals information about the names that appear in the LibriSpeech train set. Then in a real-world application, the model could leak similar sensitive information about its training data. 

\mypar{Model} We use the Conformer (L) architecture and the training method from \cite{gulati2020conformer} for training the baseline model on LibriSpeech. For all our experiments, we conduct our analysis on a checkpoint trained for $\sim$100k steps. 

\mypar{Noise Masking for Name Extraction}  Following the design in Section~\ref{sec:method}, for our implementation of \attname{} we start with the target knowledge that in English books, the title `mister' is often followed by names. For transcript generation, we use all the transcripts containing the above target structure from the Librispeech train and test partitions. 
    The train set contains $\sim$9.6k transcripts having the title `mister' followed by a name, whereas the test set contains \num{111} such transcripts.
    
    Next, for each transcript, we use two types of utterance generation: using actual speakers, and synthetic Text-To-Speech (TTS) voices.
    For actual speaker voices, we use the utterances from Librispeech train and test sets.  
    For synthetic generation, we use four voices (two male, two female) using a WaveNet TTS system \cite{tts} to generate TTS utterances for each transcript.
    
    Thus, we categorize our evaluation into four sets: 
\begin{itemize}[leftmargin=*]
    \item \textbf{Train:} Transcripts and utterances from LibriSpeech train set. Using this set can, for instance, allow a data practitioner to empirically test for data leakage before deploying a model.
    \item \textbf{Test:} Transcripts and utterances from LibriSpeech test set. Less restrictive since it assumes an analyst can obtain transcripts from a similar distribution as the training data, and use a disjoint set of speakers for utterance generation.\footnote{In LibriSpeech, the sets of training and test speakers are disjoint.} We defer analysis using out-of-distribution transcripts to future work.
    \item \textbf{Train TTS:} Transcripts from the train set, and utterances using synthetic TTS generation. Can be used in situations where an analyst has access to the train transcripts (without the sensitive names), but not necessarily an actual speaker.
    \item \textbf{Test TTS:} Transcripts from the test partition, and utterances using synthetic TTS generation. The least restrictive setting where an analyst only has access to some target transcripts similar to those used for training.
\end{itemize}

 For noise addition, we consider six types of noise sources: `silence', `car' cabin noise,   `cafe' chatter with background noise, `music',  `kitchen' background noise, and  `podcast'.

\begin{figure}[ht!]
\vspace{-0.2cm}
\centering
\includegraphics[width=0.45\textwidth]{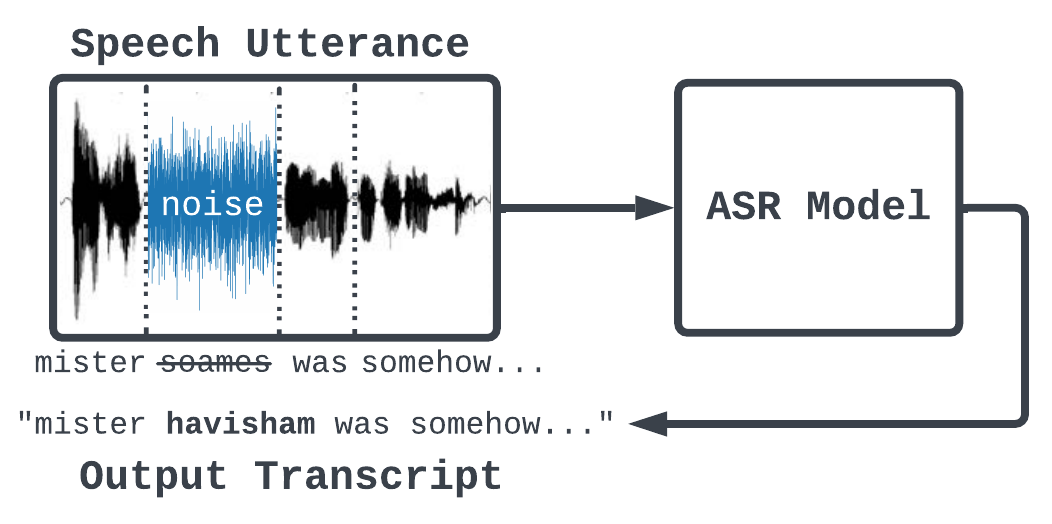}
\vspace{-0.2cm}
\caption{An illustration of \attname{} for name extraction on a LibriSpeech utterance. The  name (`soames') is replaced with pure noise in the input utterance, and fed to a model for inference. In its output transcription, the model substitutes noise with another name (`havisham') from the train set.}\label{fig:asr}
\vspace{-0.2cm}
\end{figure}

\mypar{Procedure} Our extraction method consists of passing a \emph{noise-masked} utterance (from an actual speaker, or generated via TTS) to the trained model for inference.
Every utterance contains at least one occurrence of the title `mister'.
For each utterance, we use a non-streaming Transformer-Transducer (T-T) model with RNN-T Viterbi forced alignment~\cite{KLTZS21} to calculate the time-alignment of words, and we replace the audio for the word appearing after `mister' with pure noise. For each masked name, we also mask an additional $100$ ms on both sides of it as a margin to avoid any residual information.
For instance, in the utterance  ``mister \underline{soames} was somehow \ldots",
the name `soames' is replaced with pure noise (see Figure~\ref{fig:asr} for an illustration). 
Though the duration of the noise we add here uses the duration of the masked word in the utterance, our method is also applicable to arbitrary noise lengths. We provide some results for some fixed noise duration (set without any knowledge about the duration of the masked word audio) in Section~\ref{sec:dets}.

\mypar{Performance Metrics} Our goal is to measure how often the model replaces the missing (masked) names with some name from the train set. 
To evaluate the success of our extraction method, we consider the following metrics: 
\begin{itemize}[leftmargin=*]
    \item \textbf{Average true name accuracy:} 
    This metric measures how often the model outputs the `true' name (i.e., the  name in the original utterance) from the noise-masked utterance.
    This corresponds to our highest level of leakage, and the model might have \emph{memorized} the true  name.  
    \item \textbf{Average any name success rate:}
    A slightly milder case of the above is when the model outputs \emph{some}  name from the train set. We measure the average success rate of outputting any name given noise-masked utterances.
    To construct a set of names, we tabulate the words that appear after various title words in the training dataset.
\footnote{We filter out instances containing common words, pronouns, etc., for example, `\ldots mister \underline{and} \ldots', `\ldots miss \underline{my} \ldots', etc. While our manual filtering removes many false positives for names, we expect a few edge cases to pass our filtering process, which we consider negligible for our analysis.} 
\end{itemize}

\mypar{Results} In Table~\ref{tbl:all_splits} (\#1), we present the results of \attname{}, using `silence' as well as the other five noise types, in the four different data settings (original/TTS version of train/test set) on the baseline model.
We observe a significant amount of information leakage from the baseline model in all cases, even using `silence' which is perhaps the most basic form of noise. 
\attname{} provides the highest leakage on the train utterances (e.g., 11.8\% correct name accuracy and 55.2\% any name success rate using silence). 
Remarkably, our method is able to extract significant information from TTS version of the train set (knowledge of only train transcripts), as well as both original and TTS versions of test set. 
In summary, the results indicate that the train transcript information and utterances from real speakers are advantageous, but not necessary for successful extractions using \attname{}. Even unseen (test) transcripts and TTS utterances from them can induce significant leakage.

\section{Towards Mitigations for Noise Masking}
\label{sec:mitigs}

Now, we discuss different training strategies that we empirically test towards mitigating our \attname{} method.
For each proposed strategy, we use it to train a Conformer model on LibriSpeech, and show the results of \attname{} for name extraction on the trained model in Table~\ref{tbl:all_splits}. 
We also report the Word Error Rate (WER) of each model on LibriSpeech test-clean/test-other partitions in Table~\ref{tbl:all_util}.
The performance of our baseline (\#1) from Section~\ref{sec:expts} is aligned with the Conformer benchmark on Librispeech~\cite{gulati2020conformer}.

\subsection{Upper bound: \uppname{}}
\label{sec:upp_bnd}
We start with a simple mitigation strategy designed with the knowledge of our instantiation of \attname{} (Section~\ref{sec:expts}).
A trivial way of preventing the model against such leakage is to simply remove such names from the train set before training the model.
In the \uppname{} strategy, for each train sample,
we remove all the names appearing after various title keywords (`mister', `miss', `misses', etc.) from the transcript, and 
we replace the corresponding audio in the utterance with silence.
We achieve the latter using the non-streaming T-T model as before to find the time-alignment of words.

For evaluation, we see in Table~\ref{tbl:all_util} (\#2) that the model trained using \uppname{} has utility comparable to the baseline.
From the results on \attname{} in Table~\ref{tbl:all_splits} (\#2), perhaps predictably we see that \attname{} achieves no true name accuracy in any of the considered settings. 
We observe some success (though extremely small relative to the baseline) for the any name metric.
Such leakage could be attributed to the fact that \uppname{} `scrubs' only the names appearing after certain title keywords in the dataset. 
Thus, there may still exist train samples containing names that pass such filtering.

Our results show some shortcomings of such approaches: mitigation strategies that rely on filtering targeted information can require significant domain knowledge about the data, and may still be susceptible to some leakage since the filtering is limited to a specific type of information. 
For instance, to protect users' email addresses,  we would need to apply a different filtering strategy than \uppname{}.
Note that while such a technique may not be easily scalable to all types of potentially sensitive information in a given dataset,
it essentially induces an upper-bound on the amount of mitigation we can hope to achieve in this case for models with comparable utility.

\begin{table}[t!]
\caption{Performance  of \attname{} (Average true name \% / Average any name \%) over different noises on different splits of the LibriSpeech dataset. In each case, we show the extraction result using `silence' separately from the result obtained from the rest of the noises (`Others') on average.}
\vspace{-0.2cm}
\begin{center}
\scalebox{0.75}{
\begin{tabular}{c l c c c c c} 
 \toprule
 \multirow{2}{*}{\textbf{\#}} & \multirow{2}{*}{\diagbox[width=6em]{\textbf{Model}}{\textbf{Setting}}} & \textbf{Noise} & \textbf{Train} & \textbf{Train} & \textbf{Test} & \textbf{Test} \\
 & & & & \textbf{TTS} & & \textbf{TTS} \\
 \midrule
 \multirow{2}{*}{1} & \multirow{2}{*}{Baseline} & Silence & 11.8/55.2 & 1.6/32.0 & 0.9/49.1 & 1.9/35.8 \\
  \cline{3-7}
 & & Others & 7.9/52.2 & 2.7/46.8 & 2.3/46.4 & 2.5/47.7 \\
 \cline{1-7}
 \multirow{2}{*}{2} & \multirow{2}{*}{\uppname{}} & Silence & 0.0/0.5 & 0.0/0.8 & 0.0/0.0 & 0.0/0.0 \\
 \cline{3-7}
 & & Others & 0.0/0.7 & 0.0/1.1 & 0.0/0.9 & 0.0/1.3 \\
  \cline{1-7}
  \multirow{2}{*}{3} & \multirow{2}{*}{MTR} & Silence & 12.2/49.4 & 0.7/12.8 & 0.0/36.8 & 1.9/12.3 \\
 \cline{3-7}
 & & Others & 13.5/56.6 & 5.4/50.5 & 7.4/54.2 & 4.0/48.3 \\
 \cline{1-7}
\multirow{2}{*}{4} & \multirow{2}{*}{\augname{}}  & Silence & 0.1/1.0 & 0.0/1.2 & 0.0/1.9 & 0.0/0.9 \\
 \cline{3-7} 
 & & Others & 11.9/56.4 & 5.0/53.6 & 6.2/47.5 & 4.3/45.3 \\
 \cline{1-7} 
\multirow{2}{*}{5} & \multirow{2}{*}{\begin{tabular}[c]{@{}l@{}}\augname{}\\ + MTR\end{tabular}} & Silence & 0.3/1.6 & 0.1/1.6 & 0.0/2.8 & 0.0/2.8 \\
\cline{3-7}
& & Others & 5.9/21.1 & 2.2/18.1 & 2.3/17.4 & 1.1/14.2 \\
 \bottomrule
\end{tabular}
}
\label{tbl:all_splits}
\end{center}
\end{table}

\begin{table}[t!]
\vspace{-0.2cm}
\caption{Utility for all model checkpoints used for evaluation}
\vspace{-0.6cm}
\begin{center}
\resizebox{0.37\textwidth}{!}{
\begin{tabular}{llcc}
\toprule
\multirow{2}{*}{\textbf{\#}} & \multirow{2}{6em}{\textbf{Model}} & \textbf{test-clean} & \textbf{test-other}\\
& & \textbf{WER} & \textbf{WER}\\
\midrule
1 & Baseline                                                             & 2.0                                                     & 4.5                                                     \\ \hline
2 & \uppname{}                                                         & 2.1                                                     & 4.6                                                           \\ \hline
3 & MTR                                                             & 2.0                                                    & 4.4           
 \\ \hline
4 & \augname{}                                                       & 2.1                                                     & 4.5                                              \\ \hline
5 & \augname{} + MTR & 2.1                                                     & 4.6                                                     \\ \bottomrule
\end{tabular}
}
\end{center}
\vspace{-0.7cm}
\label{tbl:all_util}
\end{table}

\subsection{Multistyle TRaining (MTR)}
Multistyle TRaining (MTR)~\cite{MTR} is a popular framework~\cite{NMSPTEH18, spec} in which a room simulator is used to combine clean audio with a variety of noises. 
We choose MTR to check if training models using utterances augmented with background noises can induce robustness against \attname{}.
For all experiments using MTR in this paper, the noises for MTR training are different from those used by our \attname{} implementation. 
Following \cite{spec}, we mix clean and MTR data with an 8:2 ratio.

We see from Table~\ref{tbl:all_util} (\#3) that the model trained using MTR provides better utility than the baseline.
However, from Table~\ref{tbl:all_splits} (\#3) we see that for all of the evaluated settings, \attname{} using noises other than silence results in more prominent leakage than the baseline, especially for true name accuracy.

\subsection{\augname{}}

Building on our intuition from \uppname{}, we design a simpler approach called \augname{} where for each sample we randomly remove words from the input transcript, and replace the corresponding audio in the utterance with silence. 
\augname{} is designed to be general, and can be applied to any dataset without requiring much domain knowledge about it.
For the experiments in this paper using \augname{}, we mask one word randomly from each utterance.
However, the number of masked words can be treated as a hyperparameter in general. 

Table~\ref{tbl:all_util} (\#4) shows that the model trained using \augname{} provides comparable utility as the baseline.
Moreover, we can see from Table~\ref{tbl:all_splits} (\#4) that while \augname{} provides extreme robustness against \attname{} with `silence', we observe no mitigation for the other noises considered.
 
Next, we combine \augname{} and MTR to see if training using utterances with randomly silenced words and augmented background noises  can provide mitigation against \attname{}. 
We see from Table~\ref{tbl:all_util} (\#5) that the model trained using \augname{} + MTR provides a comparable utility as the baseline.
Moreover, we see from Table~\ref{tbl:all_splits} (\#5) that our combined strategy achieves significant robustness against \attname{} in all four settings for all types of noises considered (detailed results in Section~\ref{sec:dets}).
In each case, the leakage via \attname{} is significantly lower than the baseline, especially for the any name success metric.
 Thus, we see that while using MTR by itself can result in an increased leakage via \attname{}, and \augname{} alone only provides mitigation against \attname{} using `silence', their combination provides substantial mitigation against a variety of noises while maintaining comparable model utility as the baseline.
 It is important to note that unlike \uppname{}, \augname{} can be applied without any domain knowledge about a dataset. 

\mypar{Note} The technique of Differentially Private Stochastic Gradient Descent
(DP-SGD)~\cite{BST14, DP-DL} has been often used for obtaining guaranteed bounds on the amount of leakage about the training data from the training process.
However, extending the bounds to target types of training data (e.g., via a group privacy argument~\cite{dwork2014algorithmic}) can result in weak guarantees.
Moreover, we consider strategies using which models can provide comparable utility as the baseline without increasing the computation cost significantly.
Since the privacy-utility-computation trade-offs for DP-SGD can be substantial for large models \cite{BST14, kairouz2021practical}, we do not provide a comparison using DP-SGD.

\section{Detailed Analysis for Leakage}
\label{sec:dets}
We provide a detailed analysis of our \attname{}  results above using LibriSpeech train utterances, aimed at helping practitioners check for such leakage before deploying a model. Specifically, we design two additional metrics, and show results for all the six noises used for \attname{} on each model.

\mypar{Additional Metrics} The following two metrics help further understand the information leakage for each case. 
\begin{itemize}[leftmargin=*]
    \item \textbf{Number of unique names:} There are \num{3622} unique names in the filtered set of names we use to compute the any name success rate. 
    We measure the total number of unique names among the names extracted from a model. 
     \item \textbf{Number of extrapolated names:} While we only use `mister' for our extraction results, we measure how many of the extracted names do not ever appear after `mister' in the train set.
    Such leakages could have resulted from a model's extrapolation of names coming after other title keywords. 
    Among the \num{3622} unique names, \num{1084} never appear after `mister'.
\end{itemize}

\mypar{Detailed Results on the Train Set} 
In Table~\ref{tbl:mtr}, we show the results using different types of noises on the train set.
We see that regardless of the noise type, there is a high chance the baseline outputs a  name given a noise masked utterance as the input. 
Moreover, we see that the baseline is able to extrapolate for a lot of names not strictly following the target structure. 
For instance, using `Car' noise, we can recover 447 out of \num{3622} unique names, 42 of which never even appear after `mister' in the train set. 
The model trained with \augname{} + MTR is more robust than the baseline in terms of all the considered metrics across all the noises. 
\ifsupp
In Appendix~\ref{sec:app_expts}, we provide the detailed results for all the models considered in this paper, on all the four evaluation settings. 
\fi

\begin{table}[ht!]
\caption{Detailed results of our \attname{} implementation using train utterances and different types of noise.}
\vspace{-0.5cm}
\begin{center}
\scalebox{0.85}{
\begin{tabular}{c r c c c c } 
 \toprule
\textbf{Model} & \textbf{Noise} & \textbf{True} &  \textbf{Any} & \textbf{Unique} & \textbf{Extrapolated}\\ 
 \midrule
 \multirow{6}{4em}{Baseline}  & Silence & 11.8\% & 55.2\% & 457 & 41 \\
\cline{3-6}
&  Car & 7.3\% & 47.7\% & 447 & 42 \\
\cline{3-6}
&  Cafe & 7.9\% & 50.4\% &  478 & 38 \\
\cline{3-6}
&  Music & 9.0\% & 62.1\% &  539 & 51 \\
\cline{3-6}
&  Kitchen & 7.7\% & 47.2\% &  463 & 50 \\
\cline{3-6}
&  Podcast & 7.5\% & 53.9\% & 595 & 73 \\
\midrule
 \multirow{6}{4em}{\augname{} + MTR} & Silence & 0.3\% & 1.6\% &  101 & 5 \\
\cline{3-6}
& Car & 6.1\% & 19.5\% &  358 & 22 \\
\cline{3-6}
& Cafe & 5.9\% & 19.2\% &  359 & 24 \\
\cline{3-6}
& Music & 5.9\% & 20.6\% &  405 & 24 \\
\cline{3-6}
& Kitchen & 6.1\% & 19.2\% &  353 & 22 \\
\cline{3-6}
& Podcast & 5.6\% & 26.9\% &  558 & 68 \\
 \bottomrule
\end{tabular}
}
\label{tbl:mtr}
\end{center}
\vspace{-0.3cm}
\end{table}

\begin{table}[ht!]
\caption{Effect of noise duration on the extraction results using `silence' and train utterances on the baseline model. `Original' corresponds to the setting where the duration of noise added uses the duration of the masked word in the utterance.}
\vspace{-0.5cm}
\begin{center}
\scalebox{0.85}{
\begin{tabular}{l c c c c} 
 \toprule
\textbf{Duration} & \textbf{True} &  \textbf{Any} & \textbf{Unique} & \textbf{Extrapolated}\\ 
 \midrule
 Original & 11.8\% & 55.2\% & 457 & 41  \\
\hline
100 ms & 6.3\% & 30.1\% &  421 & 40  \\
200 ms & 6.3\% & 30.7\% & 453 & 42  \\
500 ms & 6.2\% & 29.4\% &  430 & 35  \\
800 ms & 6.0\% & 29.5\% &  432 & 43  \\
1000 ms & 6.5\% & 30.3\% &  433 & 35  \\
1500 ms & 6.3\% & 29.9\% &  433 & 44  \\
 
 \bottomrule
\end{tabular}
}
\label{tbl:var}
\end{center}
\vspace{-0.65cm}
\end{table}

\mypar{Results with Fixed Noise Duration} In Table~\ref{tbl:var}, we show the results of \attname{} using train utterances and a fixed duration (set with no knowledge of the masked word duration) of `silence' noise on the baseline model.\footnote{We also conduct experiments for the other five noises on the baseline, and the results show similar trends as shown in Table~\ref{tbl:var}. \ifsupp We present the rest of the results in Appendix~\ref{sec:app_expts}.\else We omit the other results due to space constraints.\fi} For reference, we provide the results with noise using the `original' masked word duration. 
We see that while using the masked word duration provides a boost to the true/any name metrics, a fixed noise duration results in essentially the same amount of unique/extrapolated names being extracted across the range of durations used.

\section{Conclusion}
\label{sec:conc}

In this work, we designed \attname{}, the first method to our knowledge for demonstrating leakage of training data from trained ASR models.
We conducted experiments in four settings to show the success of our method for extracting names from LibriSpeech used for training a SOTA Conformer model.
Lastly, we designed \augname{}, a data augmentation method that, when combined with MTR,  provides comparable utility as the baseline along with significantly mitigating extraction via \attname{} in all evaluated settings.

It is important to note that the instantiation of our \attname{} method used in this paper is intended to illustrate a specific type of information leakage from a trained model.
Devising noises that aid extraction better can only improve the performance of our method.
Moreover, there can be a variety of target knowledge with which \attname{} can be used for other sensitive extractions, and it can be practically infeasible to design mitigation methods customized to each type of extraction (e.g., like we designed the \uppname{} strategy in Section~\ref{sec:upp_bnd}).
While we designed \augname{} with general applicability in mind, approaches for training ASR models that can provide better protection against leakages of structured training data while achieving utility comparable to SOTA models is an interesting direction that we leave for future work.

\bibliographystyle{IEEEtran}
\bibliography{references}

\ifsupp
\appendix
\onecolumn
\section{Additional Experimental Results}
\label{sec:app_expts}
In Tables~\ref{tbl:train_full}-\ref{tbl:test_tts_full}, we provide the detailed results for each setting (train, train TTS, test, and test TTS utterances) and using each type of noise for \attname{} on all the models evaluated in this paper. 
Specifically, we demonstrate the results for the baseline, Name Silencing, MTR, \augname{}, and \augname{} + MTR models. 

\begin{table}[ht!]
\caption{Detailed results of our \attname{} implementation using train utterances and different types of noise.}
\vspace{-0.5cm}
\begin{center}
\scalebox{0.75}{
\begin{tabular}{c r c c c c } 
 \toprule
\textbf{Model} & \textbf{Noise} & \textbf{True} &  \textbf{Any} & \textbf{Unique} & \textbf{Extrapolated}\\ 
 \midrule
 \multirow{6}{4em}{Baseline}  & Silence & 11.8\% & 55.2\% & 457 & 41 \\
\cline{3-6}
& Car & 7.3\% & 47.7\% & 447 & 42 \\
\cline{3-6}
& Cafe & 7.9\% & 50.4\% & 478 & 38 \\
\cline{3-6}
& Music & 9.0\% & 62.1\% & 539 & 51 \\
\cline{3-6}
& Kitchen & 7.7\% & 47.2\% & 463 & 50 \\
\cline{3-6}
& Podcast & 7.5\% & 53.9\% & 595 & 73 \\
\midrule
\multirow{6}{4em}{Name Silencing} & Silence & 0.0\% & 0.5\% & 33 & 4 \\
\cline{3-6}
& Car & 0.0\% & 0.7\% & 56 & 5 \\
\cline{3-6}
& Cafe & 0.0\% & 0.5\% & 40 & 5 \\
\cline{3-6}
& Music & 0.0\% & 0.7\% & 58 & 6 \\
\cline{3-6}
& Kitchen & 0.0\% & 0.6\% & 55 & 5 \\
\cline{3-6}
& Podcast & 0.0\% & 1.2\% & 75 & 14 \\
 \midrule
 \multirow{6}{4em}{MTR}  & Silence & 12.2\% & 49.4\% & 422 & 35 \\
\cline{3-6}
& Car & 12.2\% & 48.9\% & 503 & 37 \\
\cline{3-6}
& Cafe & 12.7\% & 49.7\% & 499 & 36 \\
\cline{3-6}
& Music & 16.1\% & 73.5\% & 588 & 48 \\
\cline{3-6}
& Kitchen & 12.5\% & 49.7\% & 510 & 46 \\
\cline{3-6}
& Podcast & 14.0\% & 61.2\% & 646 & 77 \\
 \midrule
 \multirow{6}{4em}{\augname{}} & Silence & 0.1\% & 1.0\% & 70 & 5 \\
\cline{3-6}
& Car & 10.5\% & 48.3\% & 456 & 34 \\
\cline{3-6}
& Cafe & 10.9\% & 48.6\% & 464 & 33 \\
\cline{3-6}
& Music & 17.0\% & 84.6\% & 584 & 42 \\
\cline{3-6}
& Kitchen & 10.5\% & 46.6\% & 451 & 34 \\
\cline{3-6}
& Podcast & 10.8\% & 54.1\% & 610 & 69 \\
 \midrule
 \multirow{6}{4em}{\augname{} + MTR} & Silence & 0.3\% & 1.6\% &  101 & 5 \\
\cline{3-6}
& Car & 6.1\% & 19.5\% &  358 & 22 \\
\cline{3-6}
& Cafe & 5.9\% & 19.2\% &  359 & 24 \\
\cline{3-6}
& Music & 5.9\% & 20.6\% &  405 & 24 \\
\cline{3-6}
& Kitchen & 6.1\% & 19.2\% &  353 & 22 \\
\cline{3-6}
& Podcast & 5.6\% & 26.9\% &  558 & 68 \\
 \bottomrule
\end{tabular}
}
\label{tbl:train_full}
\end{center}
\end{table}

\begin{table}[ht!]
\vspace{-0.5cm}
\caption{Detailed results of our \attname{} implementation using train TTS utterances and different types of noise.}
\vspace{-0.5cm}
\begin{center}
\scalebox{0.75}{
\begin{tabular}{c r c c c c } 
 \toprule
\textbf{Model} & \textbf{Noise} & \textbf{True} &  \textbf{Any} & \textbf{Unique} & \textbf{Extrapolated}\\ 
 \midrule
 \multirow{6}{4em}{Baseline} & Silence & 1.6\% & 32.0\% & 284 & 32 \\
\cline{3-6}
& Car & 2.5\% & 45.5\% & 450 & 55 \\
\cline{3-6}
& Cafe & 2.6\% & 45.0\% & 432 & 49 \\
\cline{3-6}
& Music & 3.0\% & 49.4\% & 514 & 67 \\
\cline{3-6}
& Kitchen & 2.7\% & 45.1\% & 439 & 47 \\
\cline{3-6}
& Podcast & 2.7\% & 48.9\% & 561 & 72 \\
\midrule
\multirow{6}{4em}{Name Silencing} & Silence & 0.0\% & 0.8\% & 55 & 9 \\
\cline{3-6}
& Car & 0.0\% & 1.0\% & 124 & 23 \\
\cline{3-6}
& Cafe & 0.0\% & 1.0\% & 113 & 21 \\
\cline{3-6}
& Music & 0.0\% & 1.2\% & 138 & 26 \\
\cline{3-6}
& Kitchen & 0.0\% & 1.0\% & 110 & 23 \\
\cline{3-6}
& Podcast & 0.0\% & 1.2\% & 144 & 25 \\
\midrule
\multirow{6}{4em}{MTR} & Silence & 0.7\% & 12.8\% & 226 & 33 \\
\cline{3-6}
& Car & 5.0\% & 46.2\% & 540 & 56 \\
\cline{3-6}
& Cafe & 5.2\% & 46.4\% & 560 & 56 \\
\cline{3-6}
& Music & 6.0\% & 56.0\% & 615 & 61 \\
\cline{3-6}
& Kitchen & 5.1\% & 46.8\% & 552 & 57 \\
\cline{3-6}
& Podcast & 5.7\% & 57.3\% & 720 & 90 \\
\midrule
\multirow{6}{4em}{\augname{}} & Silence & 0.0\% & 1.2\% & 85 & 15 \\
\cline{3-6}
& Car & 4.2\% & 45.2\% & 465 & 44 \\
\cline{3-6}
& Cafe & 4.6\% & 45.5\% & 497 & 47 \\
\cline{3-6}
& Music & 7.2\% & 80.7\% & 588 & 63 \\
\cline{3-6}
& Kitchen & 4.3\% & 44.9\% & 480 & 53 \\
\cline{3-6}
& Podcast & 4.7\% & 51.5\% & 623 & 81 \\
\midrule
\multirow{6}{4em}{\augname{} + MTR} & Silence & 0.1\% & 1.6\% & 115 & 22 \\
\cline{3-6}
& Car & 2.3\% & 18.4\% & 446 & 47 \\
\cline{3-6}
& Cafe & 2.3\% & 17.4\% & 452 & 50 \\
\cline{3-6}
& Music & 2.0\% & 16.7\% & 468 & 50 \\
\cline{3-6}
& Kitchen & 2.3\% & 17.6\% & 423 & 39 \\
\cline{3-6}
& Podcast & 1.9\% & 20.4\% & 636 & 89 \\
 \bottomrule
\end{tabular}
}
\label{tbl:train_tts_full}
\end{center}
\end{table}

\begin{table}[ht!]
\caption{Detailed results of our \attname{} implementation using test utterances and different types of noise.}
\vspace{-0.5cm}
\begin{center}
\scalebox{0.75}{
\begin{tabular}{c r c c c c } 
 \toprule
\textbf{Model} & \textbf{Noise} & \textbf{True} &  \textbf{Any} & \textbf{Unique} & \textbf{Extrapolated}\\ 
 \midrule
\multirow{6}{4em}{Baseline} & Silence & 0.9\% & 49.1\% & 26 & 1 \\
\cline{3-6}
& Car & 3.8\% & 44.3\% & 28 & 0 \\
\cline{3-6}
& Cafe & 0.0\% & 41.5\% & 26 & 1 \\
\cline{3-6}
& Music & 3.8\% & 54.7\% & 36 & 0 \\
\cline{3-6}
& Kitchen & 1.9\% & 41.5\% & 29 & 1 \\
\cline{3-6}
& Podcast & 1.9\% & 50.0\% & 34 & 2 \\
\midrule
\multirow{6}{4em}{Name Silencing} & Silence & 0.0\% & 0.0\% & 0 & 0 \\
\cline{3-6}
& Car & 0.0\% & 1.9\% & 2 & 0 \\
\cline{3-6}
& Cafe & 0.0\% & 0.9\% & 1 & 0 \\
\cline{3-6}
& Music & 0.0\% & 1.9\% & 2 & 0 \\
\cline{3-6}
& Kitchen & 0.0\% & 0.0\% & 0 & 0 \\
\cline{3-6}
& Podcast & 0.0\% & 0.0\% & 0 & 0 \\
\midrule
\multirow{6}{4em}{MTR} & Silence & 0.0\% & 36.8\% & 14 & 0 \\
\cline{3-6}
& Car & 7.5\% & 51.9\% & 36 & 0 \\
\cline{3-6}
& Cafe & 6.6\% & 44.3\% & 31 & 0 \\
\cline{3-6}
& Music & 6.6\% & 62.3\% & 44 & 1 \\
\cline{3-6}
& Kitchen & 7.5\% & 57.5\% & 36 & 0 \\
\cline{3-6}
& Podcast & 8.5\% & 54.7\% & 38 & 2 \\
\midrule
\multirow{6}{4em}{\augname{}} & Silence & 0.0\% & 1.9\% & 2 & 0 \\
\cline{3-6}
& Car & 6.6\% & 42.5\% & 34 & 2 \\
\cline{3-6}
& Cafe & 7.5\% & 40.6\% & 25 & 1 \\
\cline{3-6}
& Music & 7.5\% & 71.7\% & 39 & 1 \\
\cline{3-6}
& Kitchen & 4.7\% & 39.6\% & 32 & 1 \\
\cline{3-6}
& Podcast & 4.7\% & 43.4\% & 38 & 2 \\
\midrule
\multirow{6}{4em}{\augname{} + MTR} & Silence & 0.0\% & 2.8\% & 3 & 1 \\
\cline{3-6}
& Car & 2.8\% & 11.3\% & 10 & 1 \\
\cline{3-6}
& Cafe & 2.8\% & 18.9\% & 14 & 0 \\
\cline{3-6}
& Music & 0.9\% & 15.1\% & 16 & 1 \\
\cline{3-6}
& Kitchen & 2.8\% & 17.9\% & 16 & 1 \\
\cline{3-6}
& Podcast & 1.9\% & 23.6\% & 23 & 2 \\
 \bottomrule
\end{tabular}
}
\label{tbl:test}
\end{center}
\end{table}

\begin{table}[ht!]
\vspace{-0.5cm}
\caption{Detailed results of our \attname{} implementation using test TTS utterances and different types of noise.}
\vspace{-0.5cm}
\begin{center}
\scalebox{0.75}{
\begin{tabular}{c r c c c c } 
 \toprule
\textbf{Model} & \textbf{Noise} & \textbf{True} &  \textbf{Any} & \textbf{Unique} & \textbf{Extrapolated}\\ 
 \midrule
\multirow{6}{4em}{Baseline} & Silence & 1.9\% & 35.8\% & 17 & 1 \\
\cline{3-6}
& Car & 0.9\% & 56.6\% & 28 & 1 \\
\cline{3-6}
& Cafe & 2.8\% & 46.2\% & 22 & 2 \\
\cline{3-6}
& Music & 1.9\% & 44.3\% & 29 & 0 \\
\cline{3-6}
& Kitchen & 6.6\% & 50.0\% & 25 & 0 \\
\cline{3-6}
& Podcast & 0.0\% & 41.5\% & 25 & 0 \\
\midrule
\multirow{6}{4em}{Name Silencing} & Silence & 0.0\% & 0.0\% & 0 & 0 \\
\cline{3-6}
& Car & 0.0\% & 0.9\% & 1 & 1 \\
\cline{3-6}
& Cafe & 0.0\% & 1.9\% & 2 & 0 \\
\cline{3-6}
& Music & 0.0\% & 0.0\% & 0 & 0 \\
\cline{3-6}
& Kitchen & 0.0\% & 2.8\% & 3 & 0 \\
\cline{3-6}
& Podcast & 0.0\% & 0.9\% & 1 & 0 \\
\midrule
\multirow{6}{4em}{MTR} & Silence & 1.9\% & 12.3\% & 8 & 0 \\
\cline{3-6}
& Car & 0.9\% & 42.5\% & 32 & 0 \\
\cline{3-6}
& Cafe & 3.8\% & 44.3\% & 33 & 0 \\
\cline{3-6}
& Music & 6.6\% & 52.8\% & 34 & 0 \\
\cline{3-6}
& Kitchen & 3.8\% & 45.3\% & 36 & 0 \\
\cline{3-6}
& Podcast & 4.7\% & 56.6\% & 40 & 1 \\
\midrule
\multirow{6}{4em}{\augname{}} & Silence & 0.0\% & 0.9\% & 1 & 0 \\
\cline{3-6}
& Car & 3.8\% & 39.6\% & 26 & 0 \\
\cline{3-6}
& Cafe & 4.7\% & 37.7\% & 25 & 0 \\
\cline{3-6}
& Music & 7.5\% & 68.9\% & 42 & 1 \\
\cline{3-6}
& Kitchen & 2.8\% & 34.0\% & 26 & 0 \\
\cline{3-6}
& Podcast & 2.8\% & 46.2\% & 33 & 1 \\
\midrule
\multirow{6}{4em}{\augname{} + MTR} & Silence & 0.0\% & 2.8\% & 3 & 2 \\
\cline{3-6}
& Car & 0.9\% & 14.2\% & 12 & 0 \\
\cline{3-6}
& Cafe & 2.8\% & 19.8\% & 16 & 1 \\
\cline{3-6}
& Music & 0.9\% & 10.4\% & 10 & 1 \\
\cline{3-6}
& Kitchen & 0.0\% & 14.2\% & 13 & 0 \\
\cline{3-6}
& Podcast & 0.9\% & 12.3\% & 13 & 1 \\
 \bottomrule
\end{tabular}
}
\label{tbl:test_tts_full}
\end{center}
\end{table}

In Table~\ref{tbl:var_len}, we also provide the results of \attname{} on the baseline model using a fixed-length noise of 100 ms duration on train utterances.
For reference, we also provide results for noise that is of the same duration as the masked word in the utterance.
We observe that although the extraction performance degrades without using the original word duration, we can still recover a significant amount of unique and extrapolated names using any type of noise.

\begin{table}[ht!]
\caption{Extraction results using different noises and train utterances on the baseline model. `Original' corresponds to the setting where the duration of noise added uses the duration of the masked word in the utterance, and `Fixed' denotes the results using a fixed-length noise of 100 ms duration.}
\vspace{-0.5cm}
\begin{center}
\scalebox{0.85}{
\begin{tabular}{c r c c c c } 
 \toprule
\textbf{Noise} & \textbf{Duration} & \textbf{True} &  \textbf{Any} & \textbf{Unique} & \textbf{Extrapolated}\\ 
 \midrule
   \multirow{2}{4em}{Silence} & Original & 11.8\% & 55.2\% & 457 & 41 \\
& Fixed & 6.3\% & 30.1\% &  421 & 40 \\
\cline{2-6}
\multirow{2}{4em}{Car} & Original & 7.3\% & 47.7\% & 447 & 42 \\
& Fixed & 0.9\% & 6.6\% & 278 & 38 \\
\cline{2-6}
\multirow{2}{4em}{Cafe} & Original & 7.9\% & 50.4\% & 478 & 38 \\
& Fixed & 1.2\% & 8.4\% & 297 & 30\\
\cline{2-6}
\multirow{2}{4em}{Music} & Original & 9.0\% & 62.1\% & 539 & 51 \\
& Fixed & 5.3\% & 40.2\% & 541 & 56\\
\cline{2-6}
\multirow{2}{4em}{Kitchen} & Original & 7.7\% & 47.2\% & 463 & 50 \\
& Fixed & 0.4\% & 4.3\% & 253 & 30\\
\cline{2-6}
\multirow{2}{4em}{Podcast} & Original & 7.5\% & 53.9\% & 595 & 73 \\
& Fixed & 0.9\% & 12.4\% & 462 & 57\\
 \bottomrule
\end{tabular}
}
\label{tbl:var_len}
\end{center}
\end{table}
\fi
\end{document}